\documentclass[aps,pre,onecolumn,epsf,graphicx,a4]{revtex4}
\usepackage[dvips]{graphicx}
\usepackage{amsfonts}
\usepackage{epsfig}
\usepackage{amsmath}
\usepackage{amssymb}

\def\pd2x{{\partial^2 \over \partial x^2}}

\usepackage{epsfig,latexsym}

\newcommand \bew {\begin{widetext}}
\newcommand \enw {\end{widetext}}

\begin{document}

\title{\bf\noindent The statistical mechanics of combinatorial optimization 
problems with site disorder} 

\author{David S. Dean$^{(1)}$, David Lancaster$^{(2)}$ 
and Satya.N. Majumdar$^{(3)}$}

\affiliation{
(1) Laboratoire de Physique Th\'eorique,  UMR CNRS 5152, IRSAMC, Universit\'e 
Paul Sabatier, 118 route de Narbonne, 31062 Toulouse Cedex 04, France\\
(2) Harrow School of Computer Science, University of Westminster, 
Harrow, HA1 3TP, UK \\
 (3) Laboratoire de Physique Th\'eorique et Mod\`eles Statistiques, UMR 8626,
Universit\'e Paris Sud, B\^at 100, 91045 Orsay Cedex, France
}
\date{18 April 2005}
\begin{abstract}
We study the statistical mechanics of a class of problems whose
phase space is the set of permutations of an ensemble of quenched
random positions. Specific examples analyzed are the 
finite temperature traveling salesman problem on several different domains
and various problems in one dimension
such as the so called descent problem. 
We first motivate our method by analyzing these problems 
using the annealed approximation, then the limit of a large number of 
points we develop a formalism to carry out the quenched calculation. This
formalism does not require the replica method and its predictions 
are found to agree with Monte
Carlo simulations. In addition our method reproduces an exact mathematical
result for the Maximum traveling salesman problem in two dimensions and
suggests its generalization to higher dimensions.  The general approach 
may provide an alternative method to study certain systems with quenched disorder. 
\end{abstract}

\maketitle
\vspace{.2cm} \pagenumbering{arabic}

\section{Introduction}
The statistical mechanical approach to the study of optimization problems
has lead to progress in a number of ways. The approach is based
on identifying the cost function, which needs to be minimized, with
the energy of a physical system whose phase space is equivalent to the 
free adjustable parameters in the optimization problem. The zero
temperature energy of the resulting physical system thus corresponds
to the optimal solution. This formulation can be exploited in two ways.
First, physically motivated minimization techniques 
such as simulated annealing can be applied to 
optimization problems \cite{siman}, often leading to near optimal solutions.
Secondly the statistical mechanical approach can also be used to 
carry out computations of average or typical values of 
optimal solutions, where the non-adjustable 
parameters (describing the realization of the instance)
in the  system are taken to be quenched random variables
\cite{mamoze}. 
The replica and cavity methods, which are 
much used in the theory of spin glasses,
have been successfully exploited to study  statistical properties in  
wide range of optimization problems \cite{mamoze,mepavi,fuan,match}. 
Often optimization problems have a phase
space which is equivalent to permutations or partitions of the integers
and these problems are referred to as combinatorial optimization problems.
One of the most famous of these combinatorial problems is the 
traveling salesman problem (TSP). Here the problem is to find the 
minimal circuit length to visit $N$ cities or points  where the distance 
between the points  $i$ and $j$ is given by $d_{ij}$. 
The order in 
which the cites are visited is encoded in a permutation $\sigma\in \Sigma_N$
where $\Sigma_N$ is the group of permutations of $N$ objects. For a given
permutation 
\begin{equation}
D(\sigma) = \sum_i d_{\sigma_i,\sigma_{i+1}},
\end{equation}
is the corresponding total distance traveled. When the $d_{ij}$s are 
chosen from some quenched distribution the problem is referred to as the 
stochastic TSP.
The most natural form of the TSP is the Euclidean TSP \cite{tspe}
where the cities are points ${{\bf r}_1,{\bf r}_2\cdots {\bf r}_N}$ in some 
connected domain ${\cal D}$ in $\mathbb{R}^d$ and each point is 
independently distributed from the others with the same probability 
density function $p_q({\bf r})$. The distance between the points 
$i$ and $j$ is 
simply the Euclidean distance on $\mathbb{R}^d$ given by $d_{ij} = |{\bf r}_i-{\bf r}_j|$. 
It was shown \cite{tspe} that for $N\to \infty$ the  minimal path 
$D_M$ behaves as 
\begin{equation}
{D_M\over N^{1-{1\over d}}} \to \beta(d) \int_{\cal D} d^d r\ \left(p_q({\bf r}
\right))^{1-{1\over d}} \label{eqham}
\end{equation}
with probability one. Here $\beta(d)$ is a constant depending on the 
dimension of the space $d$ but independent of $p_q$. The stochastic
TSP has also been studied under the {\em random link} hypothesis
where the $d_{ij}$ are all  uncorrelated (up to any symmetry requirement).
Clearly in this random link version the triangle inequality is not respected.
This version has  been intensively studied 
\cite{tspps1,tspps2,tspcav,pema1,pema2} 
and its analysis is greatly simplified by the lack of correlation between 
the $d_{ij}$ which makes the taking of the disorder average  quite 
straight forward. A, somewhat perverse, variant of the TSP is one 
where one asks for the maximal tour, this is called the Maximum TSP 
\cite{mtsp} and is, for obvious reasons, sometimes referred to as the 
taxicab rip off.  In the statistical mechanical formulation if one 
looks for the maximal tour, one keeps the same cost function but changes the
sign of the temperature.      
 
In the class of problems we shall study in this paper, 
$N$ points $\{{\bf r}_1,{\bf r}_2\cdots,{\bf r}_N\}$
are chosen independently  in some domain ${\cal D \in  \mathbb{R}}^d$  
with probability density $p_q({\bf r})$.
Again the  dynamical phase space for the problem is taken to be all 
permutations of the order of these points, $\sigma \in \Sigma_N$. 
The Hamiltonian for the system is defined to be
\begin{equation}
H(\sigma) = \sum_{i=1}^{N} V\left({\bf r}_{\sigma_{i}} - {\bf r}_{\sigma_{i-1}}\right),
\end{equation}
Cyclic boundary conditions ${\bf r}_0 = {\bf r}_N$ are imposed. In this
context the Euclidean TSP problem corresponds to the zero temperature limit
of the case where $V({\bf r}) = {\bf r}$. 

\begin{figure}
\epsfxsize=0.5\hsize \epsfbox{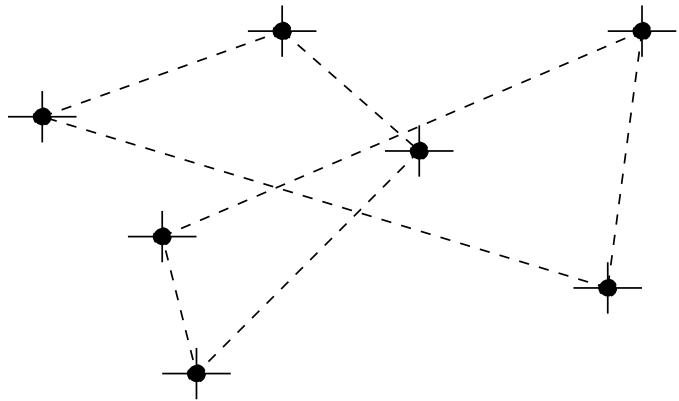}
\caption{Ring polymer on a two dimensional substrate
where each monomer (filled circles) is attached to
an impurity (crosses) }
\label{fpol}
\end{figure}

A physical realization of the system is
one where the ${\bf r}_{i}$ are impurities where the monomers of a 
polymer loop are pinned, and only one monomer can be pinned per
impurity. The potential  $V$ represents effective interaction
between neighboring monomers on the chain. 
For instance $V({\bf r}) = \lambda r^2/2$ corresponds to the Rouse model
of a polymer chain\cite{Rouse}. For instance in Fig. (\ref{fpol})
we represent the system of a polymer on a two-dimensional substrate
where the monomers, shown as filled circles, attach themselves to
pinning sites (shown as crosses).

The canonical partition function for these problems is given by 
the following sum over all permutations 
\begin{equation}
Z_N = {1\over N!}\sum_{\sigma\in {\Sigma_N}} \exp\left(-\beta H(\sigma)\right).
\label{eqpart}
\end{equation}
Since the number of permutations grows as $N!$ the entropy is 
non-extensive and behaves as $N \ln N$, but here we  insert a
factor of $1/N!$ to absorb it.

To compute the average energy per site it is necessary
to work out the quenched free energy of the system
\begin{equation}
F_N = -{1\over \beta}\overline{\ln(Z_N)},
\end{equation}
where the over-line denotes averaging with respect to the 
quenched  joint probability density function of the random sites ${\bf r_i}$.
The energy per site is then evaluated as
\begin{equation}
\epsilon = {1\over N}{\partial \over \partial \beta} \beta F_N 
\end{equation}

Our method will be shown to be  exact
in the limit of large $N$ while keeping the domain $\cal D$ fixed. However this
large $N$ scaling is not the one needed to obtain the quantity $\beta(d)$
in Eq. (\ref{eqham}). When the probability density $p_q({\bf r})$ is flat and
$V({\bf r})$ is an attractive potential, as is the case
for the ordinary TSP, the  minimal energy configuration is one where links are
always of the order of the minimal separation between points, that is to
say $O({\rho^{-{1/d}}})$ where $\rho$ is the density of points,
and thus the ground energy per site is of the order 
\begin{equation}
\epsilon_{GS} \approx V({\rho^{-{1/d}}}).
\end{equation}
If one uses the TSP potential, $V({\bf r}) = |{\bf r}|$, 
in the above, one recovers the scaling of Eq. (\ref{eqham}). 
In these attractive cases the ground state energy per site is
zero in the thermodynamic limit and in order to extract an 
extensive result the energy of the system must be scaled appropriately
with $N$ \cite{mepavi}. If the potential $V$ is repulsive, as in the Maximum 
TSP, then links will be typically of the domain size $\cal D$ and if this
domain size is $O(1)$  (in the sense that it does not scale with $N$)
then the ground state energy per site will be $O(1)$ without the need for 
any special scaling.

We emphasize that our method is exact in any dimension, however many of the 
examples we give will be in one dimension, where many  explicit
results can be obtained.In one dimension we remark that any 
attractive choice of the potential $V$ corresponds to a 
choice of cost function for the computational problem of sorting 
random data elements into increasing order.
The performance of local physical Monte Carlo algorithms has been 
analyzed in problems with these cost functions \cite{djl},
demonstrating the pitfalls in using such algorithms to search for the 
optimum. Indeed tree based sorting 
algorithms are much more efficient \cite{knuth}.

\bigskip

The main potential we shall investigate is 
\begin{equation}
V({\bf r}) = |{\bf r}|, \label{eqtspv}
\end{equation}
and, as mentioned previously, this potential is of particular interest 
as it arises naturally in the
Euclidean TSP as its ground state 
at positive temperature is the shortest circuit visiting each of
the points once and only once. 
We will discuss this problem in detail in both one and two dimensions
on several domains with different topologies. Some of these
domains are not simply connected and besides the average
energy observable, that is our main focus, we show how to compute
the statistics of the winding number of the path around the domain.
Of course the solution to the one dimensional TSP 
is obvious, one starts with the leftmost point and works
along to the rightmost, giving an average ground state energy per site 
of $\epsilon_{GS} = 0$. Using the scaling of this paper the ground state
energy of the ordinary TSP per site is  $\epsilon_{GS} = 0$ in all dimensions,
as can be seen from Eq.(\ref{eqham}). 
At negative temperature (or where the
sign of $V$ is inverted) the corresponding ground state corresponds to
the solution of the Maximum TSP where one requires the maximal distance
taken to complete a circuit visiting all the points once and only once.
The solution here is not quite so obvious but we shall see that
$\epsilon_{GS} = 1/2$ is the average ground state energy per site.
For a uniform distribution of points the interested reader may verify 
that this average value of $\epsilon_{GS}$ may be achieved by a greedy 
algorithm, which starts at the leftmost point, goes to the rightmost point,
returns to the next leftmost point and so on. We emphasize however that the 
main point of this paper is to solve the finite temperature statistical 
mechanics of these models at all temperatures.

We shall also consider the descent problem \cite{desc} as a fully soluble 
system in which all our equations can be resolved analytically. 
This is a one dimensional system  
described by the potential 
\begin{equation}
V(x) = \theta(-x). \label{eqdesv}
\end{equation}
for $x\in [0,1]$.
In fact the original formulation of the descent model, is 
equivalent to one where $x_i$ are chosen to deterministically as $x_i = i/N$.
However because of the scale free nature of the potential $V$, all  models
with an arbitrary continuous distribution of the $x_i$ are in fact
equivalent.  The energy of the permutation is thus the number of points where
$x_{\sigma_i}$ is greater than the point $x_{\sigma_{i+1}}$, the point which
follows it on the polymer ring. 
The ground state of the system is simply the permutation in which 
the $x_{\sigma_i}$ appear in increasing order and has corresponding 
energy per site $\epsilon_{GS} = 0$.

To further test the validity of our method, 
we have considered harmonic potentials $V({\bf r}) = |{\bf r}|^2$ in various
dimensions. In one dimension with domain $x_i\in [0,1]$, we have also
studied the potential $V(x) = -\ln(|x|)$ at positive temperatures (where the 
model is defined). 

In all cases studied the analytic predictions were confirmed by
Monte Carlo simulations and in some cases by extrapolating the results of  
exact enumeration for systems of
small  size. In  addition the behavior of $\epsilon_{GS}$ for repulsive 
potentials $V$ in one dimension  is analyzed via a zero 
temperature analysis, the results are discussed in terms of the corresponding 
optimal paths. We also examine the Maximum TSP in higher dimensions and
show that we recover an exact mathematical result for the average 
length of the optimal path in two dimensions. 
We are able to use our method to predict the corresponding optimal path
length in higher dimensions. A fascinating aspect of this analysis is
that in addition to providing the average optimal path length, the 
saddle point equations we derive in the thermodynamic limit seem to suggest
the heuristic one should use to search for the optimal path.

The rest of paper is organized as follows: in section (II) we study 
the annealed approximation in order to motivate the new method
for the quenched case which is presented in section (III).  Section (IV)
is concerned with the zero temperature limit and results for
the average length (or ground state energy) of optimal paths.

A brief description of our method has appeared in \cite{short} and a comment 
on the technique can be found on  {\bf http://jc-cond-mat.bel-labs.com}.

\section{The annealed approximation}
\subsection{General formalism}

Here we shall analyze the statistical mechanics of this general class
of problems in the annealed approximation which amounts to setting
\begin{equation}
F_N \approx F_N^{ann}=-{1\over \beta}\ln({\overline Z_N}),
\end{equation}
This approximation is in general doomed to failure for the following
reason. In the annealed approximation the quenched variables are
no longer quenched and will 
evolve dynamically in order to decrease the free energy of the 
system. The configuration of variables which
dominates the thermodynamics will generically  be atypical of the initial 
quenched distribution and will usually be of measure zero. For instance, if we 
consider the one-dimensional TSP at negative temperature 
it is clear that the maximal circuit, 
averaging over all permutations and positions $x_i$, will
be one where half the $x_i$ are at the point $x=0$ and the other half
at $x=1$. This will allow a ground state energy of $\epsilon_{GS}= 1$
from applying the greedy algorithm mentioned in the Introduction. However
this configuration is of measure zero if the quenched distribution is
uniform on $[0,1]$. Despite this deficiency, the formalism  below will have an
important bearing on our subsequent development of the 
quenched calculation and indeed its physical interpretation.

The partition function is averaged over all ${\bf r}_i$, $i = 1\dots N$, with 
periodic boundary conditions, {\em i.e.} ${\bf r}_0 = {\bf r}_N$. Upon this 
averaging all permutations become equivalent and we obtain
\begin{equation}
\overline{Z_N} = 
\int \prod^N_{i=1}d^dr_i\ 
\exp\left(-\beta \sum_{i=1}^{N} V({\bf r}_i-{\bf r}_{i-1})\right).
\end{equation}
The above averaged partition function can be evaluated using standard
transfer operator techniques:
\begin{equation}
\overline{Z_N} = {\rm Tr}\, T^N = \int d^dr T^N({\bf r},{\bf r}) ,
\end{equation}
where $T$ is the operator
\begin{equation}
T({\bf r},{\bf r}') = \exp\left(-\beta V({\bf r}-{\bf r}')\right).
\end{equation}
In the limit of large $N$, 
taking the size of the domain to be normalized to unity,
we find 
\begin{equation}
\overline{Z_N} = \lambda_a^N,
\end{equation}
where $\lambda_a$ is the largest eigenvalue of the operator $T$
(we use the subscript $a$ to indicate a quantity evaluated in the 
 annealed approximation).
The corresponding right and left eigenfunctions $f^{(a)}_{R,L}$ of $T$ obey
\begin{eqnarray}
f^{(a)}_R({\bf r})& = &\lambda_a^{-1}  
\int d^dr'\ \exp\left(-\beta V({\bf r}-{\bf r}')\right) f^{(a)}_R({\bf r}')\\
f^{(a)}_L({\bf r})& = &\lambda_a^{-1}  \int d^dr'\ 
\exp\left(-\beta V({\bf r}'-{\bf r})\right) f^{(a)}_L({\bf r}')
\label{eqfa}
\end{eqnarray}
These eigenfunctions are identical for a symmetric potential and we chose
the eigenfunctions to be normalized.

In this eigen-system, $\lambda_a$ can have several solutions but 
that with the maximum value of $\lambda_a$ dominates the partition
function at large $N$. Moreover the 
eigen-functions $f^{(a)}_{R,L}$ corresponding to the largest eigenvalue
must be positive by the Perron-Frobenius theorem. 
The annealed approximation for the
energy per site is consequently obtained as
\begin{equation}
\epsilon_a = -{\partial \ln(\lambda_a)\over \partial \beta}.
\label{eqea}
\end{equation}
The annealed density of points ${\bf r}_i$  on the path  at the  
point ${\bf r}$ is given by
\begin{equation}
p_a({\bf r}) = {1\over N}\langle 
\sum_{j=1}^N \delta({\bf r}-{\bf r}_j) \rangle,
\end{equation}
where the angled brackets indicates the Gibbs ensemble average 
over the annealed points. From the periodic boundary conditions, all 
points are equivalent and  we have 
\begin{eqnarray}
p_a({\bf r}) &=& \langle 
\delta({\bf r}-{\bf r}_1) \rangle = {T^N({\bf r},{\bf r})\over {\overline Z_N}}\nonumber \\
&=& f^{(a)}_R({\bf r}) f^{(a)}_L({\bf r}), \label{eqpa}
\end{eqnarray}
where again we have taken the thermodynamic limit.

In this annealed approximation the probability
density of the points  is thus given by Eq. (\ref{eqpa}).
In general we will find that
\begin{equation}
p_a({\bf r}) \neq p_q({\bf r})
\end{equation}
as the variables ${\bf r}_i$ evolve dynamically. 

\bigskip
A general expression for the energy at high temperature 
can be obtained by expanding the
averaged partition function for small $\beta$.
The first two terms of this expansion are:
\begin{equation}
\epsilon_a  = \int d^dr d^dr'\ V({\bf r}-{\bf r}')  + \beta \left[
3\left( \int d^dr d^dr'\ V({\bf r}-{\bf r}')\right)^2 - 2\int d^dr d^dr' d^dr''\ V({\bf r}-{\bf r}')V({\bf r}'-{\bf r}'')
- \int d^dr' d^dr\ V^2({\bf r}-{\bf r}') \right]. \label{eqaht}
\end{equation}
A slightly more involved calculation for the quenched case yields
the differing expansion:
\begin{equation}
\epsilon  = \int d^dr d^dr'\ V({\bf r}-{\bf r}')  + 
\beta \left[ 2 \int d^dr d^dr' d^dr''\ V({\bf r}-{\bf r}')V({\bf r}'-{\bf r}'')
- \int d^dr d^dr'\ V^2({\bf r}-{\bf r}') 
- \left( \int d^dr d^dr'\ V({\bf r}-{\bf r}')\right)^2\right].
\label{eqht}
\end{equation}
Thus, in general, there will be a difference between the annealed approximation
and quenched result at any finite temperature. We will later check our 
method for the quenched case by seeing that it reproduces the second form
Eq. (\ref{eqht}).

\subsection{The descent model}
We start by considering the annealed approximation for the descent model. 
Differentiating the first of
Eq. (\ref{eqfa}) with respect to $x$  yields 
\begin{equation}
{df^a_R\over dx} = \lambda_a^{-1} f^a_R (1-\exp(-\beta))
\end{equation}
This has the solution
\begin{equation}
f^a_R = C\exp(\kappa x) \label{eqfad}
\end{equation}
where $\kappa = \lambda_a^{-1}(1-\exp(-\beta))$ and $C$ is a constant of 
normalization. This solution is then substituted
into the original integral equation to yield 
$\lambda_a = (1-\exp(-\beta))/\beta$.
Then Eq. (\ref{eqea}) gives the annealed energy to be
\begin{equation}
\epsilon_a = {1\over \beta} - {1\over \exp(\beta)-1}\label{eqead}
\end{equation}  
The result Eq. (\ref{eqead}) is in fact identical to that obtained
from the exact solution to the descent problem, obtained via  
more lengthy combinatorial methods \cite{desc}. 
The annealed approximation thus leads to the exact energy per site
for this problem. This exactness is straightforward to understand. In this
problem there is no length scale in the potential $V$ and only the order of
the points determines the energy, clearly one would obtain the same energy
from any continuous distribution of points selected independently with
density $p_q(x)$ on $[0,1]$. We note that the solution for the other (left)
eigenvector $f^a_L$ is 
\begin{equation}
f^a_L = C'\exp(-\kappa x)
\end{equation}
and we obtain find $p_a(x) = 1 = p_q(x)$, thus the annealed distribution 
agrees with the quenched one. This is an autoconsistency  
of the annealed approximation which leads to it being exact.

\subsection{The One-Dimensional TSP}
In this section we restrict ourselves to one dimension and consider
the unit interval  with $x \in [0,1]$. 
We note that the potential $V$ is symmetric and thus may write $f_R^{(a)}=
f_L^{(a)} = f^{(a)}$. 

We proceed by differentiating Eq. (\ref{eqfa}) twice to obtain
\begin{equation}
{d^2 f^{(a)}\over dx^2} - \beta^2 f^{(a)} + 2{\beta \over\lambda_a} 
f^{(a)} = 0,
\label{eqanntsp}
\end{equation}
which has solution
\begin{equation}
f^{(a)} = C\left( \exp(\omega x) + A \exp(-\omega x)\right) \label{eqsl}
\end{equation}
with 
\begin{equation}
\lambda_a^{-1} = {\beta^2 -\omega^2 \over 2 \beta}.
\end{equation}
Now, $\lambda_a$ must be positive, so we have
\begin{eqnarray}
\omega^2 &<& \beta^2\ \ {\rm for } \ \beta > 0 \\
\omega^2 &>& \beta^2\  \ {\rm for } \ \beta < 0
\end{eqnarray}
Substituting the solution  Eq. (\ref{eqsl}) back into Eq. (\ref{eqfa})
we find the condition
\begin{equation}
\exp(2\omega) = \left({\beta -\omega\over \beta +\omega}\right)^2.
\label{eqome}
\end{equation}
We note that a solution of Eq. (\ref{eqome}) is $\omega =0$. However
the corresponding solution for $f^{(a)}$ would be of the form 
$f^{(a)}(x) = Ax +B$. From above one must also have that $df^{(a)}/dx$ 
is continuous and the clear symmetry $f^{(a)}(x) = f^{(a)}(1-x)$ means that
$A=0$ in this solution. One can verify that $f^{(a)}(x) =B$ is not a 
solution. Hence the solution must have $\omega \neq 0$. For $\beta >0$
one finds that $\omega = i\beta z$ where $z$ is the smallest positive 
solution of
\begin{equation}
  z = {\rm cot}(\beta z/2), \label{eqzp}
\end{equation}
and shows no discontinuities as $\beta$ varies.
The annealed energy per site is then given by
\begin{equation}
\epsilon_a = {1\over \beta}  - {z^2 \over \left( 1 + 
{1\over 2}\beta (1+z^2)\right)}
\end{equation}
For $\beta < 0$ we find that $\omega = \beta z$ where $z$ is the
root of 
\begin{equation}
 z = {\rm coth}(-\beta z/2).\label{eqzn}
\end{equation}
and here the annealed energy per site is
\begin{equation}
\epsilon_a = {1\over \beta}  + {z^2 \over \left( 1 + 
{1\over 2}\beta (1-z^2)\right)}
\label{eqzen}
\end{equation}
In both cases, as $\beta \to 0$ we find that
\begin{equation}
\epsilon_a(0) = {1\over 3}
\end{equation}
which agrees with the exact high temperature result Eq. (\ref{eqaht}).

In the positive temperature case the smallest positive solution $z^*$ to 
Eq. (\ref{eqzp}) is such that $z^*\in (0,\pi/\beta)$ and thus $z^*\to 0$
as $\beta \to \infty$ and thus $\epsilon_a \approx 1/\beta \to 0$; 
this clearly agrees with the limiting behavior of the corresponding 
quenched case. In the case of negative  temperature we find that 
the solution of Eq. (\ref{eqzp}) behaves as $z^* \to 1$ 
(plus an exponentially decaying correction), 
thus yielding $z^* \approx 1+ 1/\beta \to 1$ which implies
$\epsilon_a \to 1$. This latter result is clearly not the correct result 
for the quenched case as the distribution of the $x_i$ has
evolved to permit this maximal energy configuration as explained in the 
Introduction. Indeed we find from Eq. (\ref{eqsl}) that
$p_a(x) = f^2_a(x) \neq 1$ and becomes peaked at the boundaries.

\subsection{The TSP on a Ring}
When the domain of the one-dimensional TSP is a periodic ring rather than the
unit interval with boundaries considered in the last section, the analysis
becomes simpler, moreover there is an interesting new observable,
the winding number.
For a ring domain, the shortest path is sometimes the
other way round the ring, so the potential is
\begin{equation}
V(x) = 
\begin{cases}
|x|,& \ {\rm for}\  |x|<{1\over 2}  \\
1-|x|,& \  {\rm for} \ |x|>{1\over 2}
\end{cases}
\end{equation}
Taking proper account of the contributions from discontinuities in 
the derivative of this potential, the equivalent
formula to (\ref{eqanntsp}) gains a term on the right hand side:
\begin{equation}
{d^2 f^{(a)}\over dx^2} - \beta^2 f^{(a)}(x) + 2{\beta \over\lambda_a} 
f^{(a)}(x) = {2 \beta e^{-\beta/2} \over \lambda} f^{(a)}(x+1/2),
\end{equation}
where the eigenfunction is now  a periodic function with period 1.
In this case, in contrast to the situation with boundaries,
the original integral equation (\ref{eqfa}) admits a constant solution $f_a=1$.
For positive $\beta$, this must correspond to the largest
eigenvalue which is:
\begin{equation}
\lambda_a = {2(1-e^{-\beta/2})\over \beta}
\label{lambdaring}
\end{equation}
The annealed energy per site is then given by
\begin{equation}
\epsilon_a = {1\over \beta}  - 
{1\over 2(e^{\beta/2} -1)}
\label{eqring}
\end{equation}
The high temperature limit of $\epsilon_a$ now takes the value $1/4$.

Notice that the constant solution indicates that the 
annealed density of points is the same as that of the
desired quenched distribution, so the annealed
approximation is exact in this case. This result
follows from the symmetry of the domain and continues to
hold for certain other closed domains in higher dimensions.
The higher dimensional cases will be treated in the later
section on quenched models.

A new observable, the winding number, arises
for this domain. The winding number counts the number
of times a particular path goes around the ring and can be written as
the sum of contributions from each step between points
in the same way as the original Hamiltonian.
\begin{equation}
W(\sigma) = \sum_{i=1}^{N} W\left(x_{\sigma_{i}} - x_{\sigma_{i-1}}\right),
\end{equation}
Where: 
\begin{equation}
W(x) = 
\begin{cases}
1+x,& \ {\rm for}\  x<-{1\over 2}; \\
x,& \  {\rm for} \ |x|<{1\over 2};  \\
-1+x,& \  {\rm for} \ x>{1\over 2}.
\end{cases}
\end{equation}
In fact, since the cyclic boundaries lead to
$\sum (x_{\sigma_{i}} - x_{\sigma_{i-1}}) = 0$,
the $x$ part of the contributions can be dropped and the 
winding number may be written:
\begin{equation}
W(\sigma) = \sum_{i=1}^{N} 
-\theta\left(x_{\sigma_{i}} - x_{\sigma_{i-1}} - 1/2\right)
+\theta\left(-x_{\sigma_{i}} + x_{\sigma_{i-1}}- 1/2\right),
\end{equation}
In this form it shows some similarity with the potential for the descent model.
In particular, the  asymmetry will make
the right and left eigenfunctions differ.

We compute the expectation values of the winding number by
taking the Boltzmann weight of the configuration $\sigma$ to be  
\begin{equation}
\exp\left(-\beta H(\sigma) -\gamma W(\sigma)\right)
\end{equation}
The analysis
developed above holds for this slightly more general case
and we will obtain expectations of the winding number by differentiating
the partition function with respect to $\gamma$ before
setting $\gamma$ to zero.

By considering the equations obtained by differentiating
the eigenvalue equation it becomes clear that a solution
of the form:
\begin{equation}
f^a_R = C\exp(\kappa x)
\end{equation}
should be sought. Indeed, this form is a solution of the
integral equation provided $\kappa = \gamma$. As was the
case for the descent model, the left eigenvalue is
of the form $f^a_L = C'\exp(-\kappa x)$, so the
annealed density becomes constant and this solution is
thus also valid for the quenched problem.
The corresponding eigenvalue is given by 
\begin{equation}
\lambda = {2\beta\over \beta^2 - \gamma^2}
- {e^{-\beta/2}\over \beta^2 - \gamma^2}
\left((\beta+\gamma)e^{\gamma/2}
+(\beta-\gamma)e^{-\gamma/2}\right),
\end{equation}
which correctly reduces to Eq.
(\ref{lambdaring}) when $\gamma$ is set to zero.

As should be exppected, the mean value of the 
winding number vanishes since there
is nothing that prefers winding in one direction over the other. 
Even in the low temperature limit, there is no symmetry
breaking in this one-dimensional system.
On the other hand, the fluctuations provide a non
vanishing observable:
\begin{equation}
\langle \overline{ W^2} \rangle
= {N \over 4\beta^2}
{(8e^{\beta/2} - 8 - 4\beta - \beta^2) \over
(e^{\beta/2} - 1) }
\end{equation}
In the high temperature limit the fluctuations per site become $1/12$
and for large negative $\beta$ the limit is $1/4$.

\begin{figure}
\epsfxsize=0.5\hsize \epsfbox{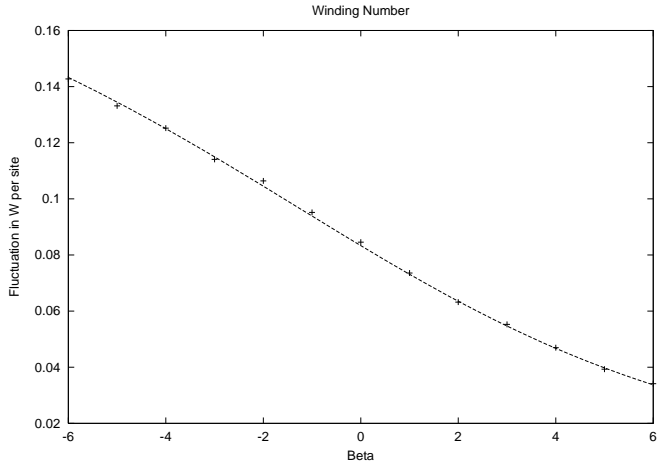}
\caption{Expectation value for the fluctuations in
winding number per site, $\langle \overline{ W^2} \rangle/N$,
for the one dimensional TSP on the ring domain as a function 
of $\beta$ (dotted line) compared with the  
Monte Carlo simulations (crosses). 
Negative $\beta$
corresponds to the Maximum TSP problem.
}
\label{fig:1DTSPWind}
\end{figure}

We have performed Monte Carlo simulations to test
this prediction using the quenched model which we have
already argued has the same value of observables. 
These results are show good agreement as displayed 
in Fig. (\ref{fig:1DTSPWind}).

\section{The quenched calculation}
\subsection{General formalism}
The order of the points ${\bf r}_i$ is unimportant for the
statistical mechanics of this problem because the phase space is 
all their possible orderings. The relevant disorder is thus clearly 
fully determined by the $N$ position vectors ${\bf r}_i$ of the sites. 
These positions are encoded in  the, unaveraged,  density of points in space
\begin{equation}
\rho_q({\bf r}) = {1\over N}\sum_{i=1}^N \delta({\bf r}-{\bf r}_i^{(q)})
\end{equation}
where we have used the superscript $q$ above to emphasize that the 
points ${\bf r}_i^{(q)}$ are quenched. We note that by definition we have
$p_q({\bf r}) ={\overline \rho_q({\bf r})}.$ and in the limit of large
$N$ we expect that
\begin{equation}
\int d^dr \ \rho_q ({\bf r}) h({\bf r}) = 
\int d^dr \ p_q ({\bf r}) h({\bf r})  +O(1/\sqrt{N}) \label{eqcl}
\end{equation}
for suitably well behaved functions $h$.
If ${\bf r}_i$ is the site visited by the polymer at step 
$i$ in a system which has ${\bf r}_i \in{\cal D}$ then the 
partition function of the permutation problem can be written as
\begin{equation}
Z_N = {1\over N!}\int \prod_{i=1}^{N} \ d^dr_i \prod_{\bf r} \left(N \rho_q({\bf r})\right)!
\prod_{\bf r}\delta\left(N \rho_q({\bf r}) - \sum_i \delta({\bf r}-{\bf r}_i)\right)
\ \exp\left(-\beta\sum_i V({\bf r}_{i+1}-{\bf r}_{i})   \right). \label{eqzq1}
\end{equation}
The above can be derived by considering a discrete version of the problem
where one has $n(j)$  sites at the points ${\bf r}(j)$. A path is 
specified by the possible sequences ${\bf r}_1, {\bf r}_2, \cdots {\bf r}_N$;
however at the  visit to the site ${\bf r}(j)$ there
are $n(j)$ possible points to chose from and thus any path has a degeneracy
$\prod_j n(j) !$ in order to have the same phase space as the permutation 
problem. In addition each site ${\bf r}(j)$ can only be visited $n(j)$ times, 
explaining the delta function constraint above. Another way of 
obtaining the factor in Eq. (\ref{eqzq1}) is to note that $Z_N(\beta) = 
C_N \Xi_N(\beta)$
where  
\begin{equation}
\Xi_N = \int \prod_{i=1}^{N} \ d^dr_i 
\prod_{\bf r}\delta\left(N \rho_q({\bf r}) - \sum_i \delta({\bf r}-{\bf r}_i)\right)
\ \exp\left(-\beta\sum_i V({\bf r}_{i+1}-{\bf r}_{i})   \right)
\end{equation}
is the partition function of the system up to a temperature independent
entropy/degeneracy contribution $C_N$. Clearly, as defined here, $Z(0) =1$ 
which  implies $Z(\beta) = \Xi(\beta)/\Xi(0)$. We shall later that, in the 
limit of large $N$ we have
\begin{equation}
\Xi_N(0) =\exp\left(-N\int d^dr \ p_q({\bf r})[\ln(p_q({\bf r}))-1]\right)
\end{equation}
which in the large $N$ limit, via Stirling's formula, recovers 
Eq. (\ref{eqzq1}). 

The partition function $\Xi_N$
may be written using a Fourier representation of the functional
constraint:
\begin{equation}
\Xi_N = \int d[\mu] \exp\left(N\int d^dr  \mu({\bf r})\rho_q({\bf r})\right)
{\cal Z}_N,
\label{eqxiq}
\end{equation}
where each integration over $\mu(x)$ is up the imaginary axis.
The object ${\cal Z}_N$ is similar to the annealed
partition function considered in the previous section,
but with an ${\bf r}$ dependent chemical potential. 
It is defined as:
\begin{equation}
{\cal Z}_N = \int \prod^N_{i=1}d^dr_i\ 
\exp\left(-\beta \sum_{i=1}^{N} V({\bf r}_i-{\bf r}_{i-1})  
-\sum_{i=0}^N \mu({\bf r}_i)\right).
\label{eqcalz}
\end{equation}
For large $N$ we may use the relation Eq. (\ref{eqcl}), neglect the terms
$O(\sqrt{N})$, and then  the partition function in Eq. (\ref{eqxiq}) 
can be evaluated by the saddle point method in the limit where 
$N\to \infty$ keeping $\cal D$ fixed. 
The saddle point equation is
\begin{equation}
p_q({\bf r}) = -{1\over N}{\delta \ln {\cal Z}_N \over \delta \mu({\bf r})}
= {1\over N}\langle\sum_{i=1}^N
\delta({\bf r}-{\bf r}_i)\rangle
 = p_a({\bf r}) 
= -{\delta \ln(\lambda_q) \over \delta \mu({\bf r})}.
\end{equation}
where the above expectation, is in the 
system with partition function ${\cal Z}_N$ defined in Eq. (\ref{eqcalz}).

Physically this approach can be thought of as choosing a site dependent 
chemical potential $\mu$ which fixes the density of the
annealed calculation to be the same as that of the quenched one
$p_q({\bf r}) = p_a({\bf r})$. This idea was used sometime ago in an 
approximative sense where  low order  moments, not the whole distribution, 
were fixed in this way \cite{morita,kuhn}. 

To proceed, we find an expression for $p_a({\bf r})$ 
using similar techniques to those employed in the last section. 
We find
\begin{equation}
{\cal Z}_N = {\rm Tr}\, {\cal T}^N
\end{equation} 
where 
\begin{equation}
{\cal T}({\bf r},{\bf r}') =\exp\left(-{\mu({\bf r})/ 2}\right)
 \exp\left(-\beta V({\bf r}-{\bf r}')\right)
\exp\left(-{\mu({\bf r}')/ 2}\right)
\end{equation} 
The  right and left ground state eigenfunction, corresponding to the 
maximal eigenvalue $\lambda_q$, obey 
\begin{eqnarray}
f^{(q)}_R({\bf r}) &=& \lambda_q^{-1} \exp\left(-{\mu({\bf r})/ 2}\right)
\int d^dr'\ \exp\left(-\beta V({\bf r}-{\bf r}')\right)
\exp\left(-{\mu({\bf r}')/ 2}\right)f^{(q)}_R({\bf r}') \\
f^{(q)}_L({\bf r}) &=& \lambda_q^{-1} \exp\left(-{\mu({\bf r})/ 2}\right)
\int d^dr'\ \exp\left(-\beta V({\bf r}'-{\bf r})\right)
\exp\left(-{\mu({\bf r}')/ 2}\right)f^{(q)}_L({\bf r}')
\label{eqfq}
\end{eqnarray}
and by a similar calculation to that of the annealed case 
we have:
\begin{equation}
p_q({\bf r}) 
= p_a({\bf r}) 
= -{\delta \ln(\lambda_q) \over \delta \mu({\bf r})}
= f^{(q)}_R({\bf r})f^{(q)}_L({\bf r}).
\end{equation}
In the case of a symmetric potential where $V({\bf r}) = V(-{\bf r})$, 
we have that
$f^{(q)}_R = f^{(q)}_L =f^{(q)} $ and thus 
$f^{(q)}= \sqrt{p_q}$, note this ensures that
$f^{(q)}$ is the  eigenfunction corresponding to
the maximal eigenvalue as we note that it is positive and 
then appeal to the Perron-Frobenius theorem. 
Substituting the above into the saddle point equation gives
\begin{equation}
\sqrt{p_q({\bf r})} = 
\lambda_q^{-1} \exp(-{\mu({\bf r})\over 2})\int d^dr'\ \exp\left(-\beta V({\bf r}-{\bf r}')\right)
\exp(-{\mu({\bf r}')\over 2})\sqrt{p_q({\bf r}')}
\end{equation}
we can thus write $\exp(-{\mu({\bf r})\over 2}) 
= \sqrt{p_q({\bf r})}/s_{\lambda_q}({\bf r})$ where
$s_{\lambda_q}({\bf r})$ obeys
\begin{equation}
s_{\lambda_q}({\bf r}) = \lambda_q^{-1} \int d^dr'\ \exp\left(-\beta V({\bf r}-{\bf r}')\right)
{pq({\bf r}') \over s_{\lambda_q}({\bf r}')} \label{eqsx}
\end{equation}
Substituting this back into the action we obtain
\begin{equation}
{\ln(Z_N)\over N} = 
2\int d^dr\ p_q({\bf r}) \ln\left(s_{\lambda_q}({\bf r})\right) \   
+ \ln(\lambda_q)
-\int d^dr \ p_q({\bf r})\ln(p_q({\bf r})) , 
\label{eqaction}
\end{equation}
the last term which is independent of $\beta$ explains the 
presence of the combinatorial term in Eq. (\ref{eqzq1}). 
However from  Eq. (\ref{eqsx}) 
we see that there is a whole family of solutions 
$\lbrace s_{\lambda_q}({\bf r}), \lambda_q\rbrace$ 
which are related by 
$s_{\lambda_q} = a^{1/2}s_{a\lambda_q}$, for $a>0$ and in addition these 
solutions all have the same action. This apparent zero mode is an artifact
introduced by the fact that the constraint 
$N\int d^dr\  \rho_q = \int d^dr\  
\sum_i\delta({\bf r}-{\bf r}_i)$ is automatically satisfied. 
Thus we may chose $\lambda_q = 1$. In the case of a uniform distribution on a domain of unit volume
this leads to our final  result
\begin{eqnarray}
\epsilon 
&=& -2{\partial \over \partial \beta}\left[ \int d^dr\ 
\ln\left(s({\bf r})\right)\right] \nonumber \\
&=& \int d^dr\ d^dr'\  {V({\bf r}-{\bf r}') 
\exp\left(-\beta V({\bf r}-{\bf r}')\right)\over
s({\bf r}) s({\bf r}')}
\label{eqsf}
\end{eqnarray}
where $s$ obeys
\begin{equation}
s({\bf r}) =\int d^dr'\ {\exp\left(-\beta V({\bf r}-{\bf r}')\right)\over 
s({\bf r}')} \label{eqsff}
\end{equation}  
and we have specialized to the usual case when the distribution
of quenched points is uniform. Recall that we have set the size of the
domain to one, and some scaling is needed when this is not the case.

It is possible to check our result via direct comparison
with the high temperature expansion Eq. (\ref{eqht}) written down at the end of section (II). Equation (\ref{eqsff}) can be solved
perturbatively as a power series in $\beta$ by writing
\begin{equation}
s({\bf r}) = 1 + \beta s_1({\bf r}) + \beta ^2 s_2({\bf r}) \cdots
\end{equation}
Substituting this expansion in to Eq. (\ref{eqsff}) yields
\begin{eqnarray}
s_1({\bf r}) &=& -\int d^dr'\ V({\bf r}-{\bf r}') + {1\over 2} \int d^dr' d^dr''\ V({\bf r}''-{\bf r}') \nonumber \\
s_2({\bf r}) &=& {1\over2} \int d^dr' \ V^2({\bf r}-{\bf r}') + \int d^dr' \ s_1({\bf r}') V({\bf r}-{\bf r}')
\label{eqss2} \\
&-&{1\over 4} \int d^dr' dz\ V^2({\bf r}'-z) -{1\over2}  \int d^dr' d^dr''\ s_1({\bf r}') V({\bf r}'-{\bf r}'')
+{1\over 2} \int d^dr' \ s_1({\bf r}')
\label{eqss1}
\end{eqnarray}
To order $\beta$ Eq. (\ref{eqsf}) then yields
\begin{equation}
\epsilon = -2 \int d^dr \ 
\left(s_1({\bf r}) + 2\beta(s_2({\bf r}) - {s_1^2({\bf r})\over 2}\right)
+ O(\beta^2)
\end{equation}
and substituting Eqs. (\ref{eqss1}) and (\ref{eqss1}) in the above we 
recover the result Eq. (\ref{eqht}).

\subsection{Quenched Rouse Polymer}
Before moving on to consider the TSP, we first analyze a
model with a harmonic potential to demonstrate an analytic
solution of the quenched equations. This system, in the 
annealed case, is used to model a polymer \cite{Rouse}.
To avoid boundaries, which prevent an analytic solution,
we treat the case where the quenched distribution of
points is Gaussian: $p_q(r)=e^{-r^2/2}/(2\pi)^{d\over2}$ and
we work in arbitrary dimension $d$.
The quenched equation for $s({\bf r})$ becomes:
\begin{equation}
s({\bf r}) =\int {d^dr'\over (2\pi)^{d\over 2}}\ { e^{-r'^2/2}
\exp\left(-\beta |{\bf r} - {\bf r}'|^2\right)\over 
s({\bf r}')}
\end{equation}  
So we search for a Gaussian solution,
\begin{equation}
s({\bf r}) = s e^{-\gamma r^2/2}
\end{equation}  
This is satisfied provided:
\begin{eqnarray}
\gamma = {1\over 2}\left( 1+2\beta -  \sqrt{1+4\beta^2}\right)\\
s = 2^{d/4}\left( 1+\sqrt{1+4\beta^2}\right)^{-d/4}
\end{eqnarray}
Inserting these into the saddle point action we obtain
the average energy per site as:
\begin{equation}
\epsilon = d\left(1- {2\beta\over\sqrt{1+4\beta^2}} + 
{2\beta\over\sqrt{1+4\beta^2}(1+\sqrt{1+4\beta^2})}\right)
\end{equation}  
This has the correct $d/2\beta$ behavior at large $\beta$
and the value $\epsilon = d$ at infinite  temperature, as can 
be checked directly by a Gaussian average.
In contrast to the situation for directed polymers, there
is no evidence for a phase transition in this quenched
case.

\begin{figure}
\epsfxsize=0.5\hsize \epsfbox{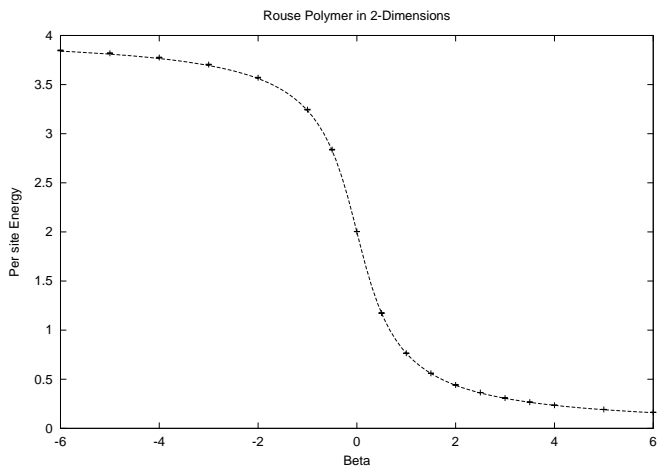}
\caption{Theoretical prediction for the average energy $\epsilon$ for 
two dimensional quenched polymer model
as a function 
of $\beta$ (solid line) compared with the  
Monte Carlo simulations (solid circles). 
Negative $\beta$
corresponds to the maximal problem.
Error bars based on 20 realizations of the quenched points
are smaller than the symbol sizes. Here the domain $\cal D$ is unbounded
but $p_q$ is Gaussian and centered at the origin}
\label{fig:Rouse2D}
\end{figure}

In figure (\ref{fig:Rouse2D}) we show the average energy
in the two-dimensional case and compare it with
Monte Carlo simulations. The agreement is good, though it
should be noted that at the edges of the plot, for large
$|\beta|$, long runs (10's of millions of steps) with large $N$ 
are required to see accurate agreement.

\subsection{TSP in One and Two Dimensions}
In this section we consider
predictions for the TSP in both one and two dimensions.
In view of the nonlinear nature of 
equations (\ref{eqsf}) and (\ref{eqsff}) 
we have not found any non-trivial analytic solutions for the TSP
potential. 
Our primary tool is
the iterative numerical solution of Eq. (\ref{eqsff})
which is stable and can be solved to any required accuracy.
However, for
closed symmetric domains (we shall consider a one dimensional ring
and a disc and torus in two dimensions),
a constant solution exists 
and some analytic progress is possible. To see this,
we simply require a domain such that that the origin of the integration in 
Eq. (\ref{eqsff}) can be shifted to yield:
\begin{equation}
s^2 =\int d^dr\ \exp\left(-\beta V({\bf r})\right) 
\label{eqconsts}
\end{equation}  
The significance of this observation is that the annealed
approximation is exact for these domains. Indeed this 
equation is exactly  the annealed eigenvalue equation for
a constant eigenfunction and
the relationship between the value of the constant $s$
and the eigenvalue $\lambda$ of the annealed approximation
is simply $\lambda = s^2$. 
In all cases the normalization is  $\lambda (\beta = 0)=1$. 
This conclusion is consistent
with what is known from analysis of the independent link
version of the TSP. Evidently, the geometry of independent links
has no boundaries and the analysis of this problem in the
same scaling limit  we consider here also shows that 
that the annealed approximation is exact \cite{tspps1}.

The expression (\ref{eqconsts}) allows a general large $\beta$
expansion for these closed domains. Provided the potential is
convex, then in this limit it is apparent that the only 
contribution to the integral is from nearby points. When the 
potential is smooth we can expand to find 
$\lambda \sim  e^{-\beta V(0)} (\beta V'')^{-d/2}$,
so the energy is $\epsilon \sim V(0) + d/2\beta$.
This expansion is invalid for the TSP (and descent model) 
since the potential
is not smooth, and the correct energy is
$\epsilon \sim d/\beta$.
These results indicate that in this limit the topology of the
domain becomes unimportant and only the dimension is relevant as
is indeed observed in all the examples below.
A similar limit for the maximal problem is dominated by a
term corresponding to the largest distance two points can be 
apart from each other, which is sensitively dependent on the
topology of the domain.

\subsubsection{One Dimension}
For the case of the ring domain, 
constant $s$ is a solution and as expected this reproduces the energy
obtained via the annealed calculation (\ref{eqring}). 

For the unit interval, we use Eq. (\ref{eqsf})
to predict the energy by  iteratively 
solving Eq. (\ref{eqsff}). 

\begin{figure}
\centering\epsfig{file=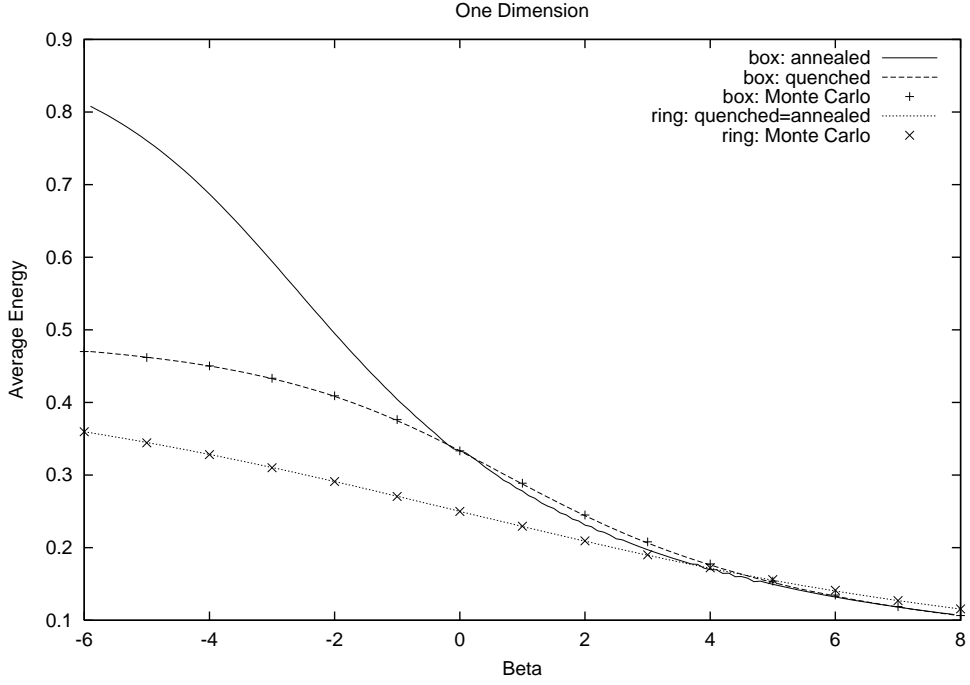,width=0.5\linewidth,angle=270}
\caption{Theoretical prediction for the average energy $\epsilon$ for 
one dimensional TSP on ring and line domains as a function 
of $\beta$ (solid line) compared with the  
Monte Carlo simulations (solid circles). 
The results of the annealed approximation for the line domain
is also shown.
Negative $\beta$
corresponds to the Maximum TSP problem.
Error bars based on 20 realizations of the quenched points
are smaller than the symbol sizes.}
\label{fig:1DTSP}
\end{figure}

To test these predictions we have carried out Monte
Carlo simulations of the TSP problem for system sizes of $N=5000$ and compared 
the average energy measured after equilibrating the system over
$5000$ Monte Carlo sweeps and measuring the average energy over a subsequent
$5000$ Monte Carlo sweeps. The disorder average was carried out by averaging 
the results over 20 independent realizations of the disorder. 
The standard move was taken to be a random 
transposition of a pair of points in the permutation and 
the acceptance of the move 
was chosen with the Metropolis rule. The results for both the
ring and line domains are shown in Fig. (\ref{fig:1DTSP})
compared with the predictions.
We see that for all temperatures the agreement is excellent.

In the same figure we also show the result for the annealed
approximation for the line
domain based on Eqs. (\ref{eqzp})-(\ref{eqzen}). We see that 
this provides a lower energy than the quenched result for
positive $\beta$ and higher for negative $\beta$.

The fluctuation in the energy 
$\langle (\epsilon - \langle \epsilon \rangle)^2\rangle$
at high temperature can be computed using the order $\beta$
term in Eq. (\ref{eqht}). This takes different values for
annealed and quenched cases, but agrees in each case
with the respective values computed using the combinatorial approach 
of reference \cite{djl}.

\subsubsection{Two Dimensions}
In two dimensions the domains we consider are the sphere,
torus and the unit square or box. 

It is convenient to treat a sphere of unit radius 
so the basic equations need a little modification to
deal with a domain $\cal D$ whose  size is not one but $\cal V$.
In effect, all integrals appearing in the formalism
are normalized by the volume, and the final expression for
the energy $E(\beta,{\cal V})$ scales as:
\begin{equation}
E(\beta, 1) = {1 \over {\cal V}^{1/d}} E(\beta/{\cal V}^{1/d}, {\cal V})
\end{equation}

The annealed/quenched equations have a constant solution with:
\begin{equation}
\lambda = s^2 = 
{1\over 4\pi} \int_{S_2} d^2r \ e^{-\beta\theta}
={1 \over 2(\beta^2 +1)} (1 + e^{-\pi\beta})
\end{equation}
leading to energy (normalized for a unit size domain)
\begin{equation}
\epsilon_a
= {2\beta \over \beta^2 +4\pi} +  {\sqrt{\pi} \over 2(e^{\sqrt{\pi}\beta/2}+1)}
\label{eqsphere}
\end{equation}
Note that in the limit $\beta \to -\infty$, the energy
becomes the  half circumference corresponding, as is the case for all
the closed domains, to the 
maximum distance two points can be apart. 

\medskip
For a torus we return to unit size domain normalization and find that
the annealed/quenched equations yield
\begin{eqnarray}
\lambda &=& s^2 = {8 \over \beta^2} \int_0^{\pi/4}
(1 - e^{-\beta/2\cos\theta} - {\beta e^{-\beta/2\cos\theta}\over 2\cos\theta}
)\,d\theta \nonumber \\
&=& {2\pi\over\beta^2} 
-{8\over\beta^2}\int_0^{\ln (1+\sqrt{2})}
({1+ {\beta\over  2}\cosh v \over \cosh v}
e^{ -{\beta\over  2}\cosh v })\ dv
\label{eqtorus}
\end{eqnarray}
In the high temperature limit the integral can be
evaluated analytically leading to:
\begin{equation}
\epsilon_a(\beta = 0)
= {1\over 6}\left( \sqrt{2} + \ln (1 + \sqrt{2})\right) 
= 0.382597858 
\end{equation}
At very low temperature we can do a saddle point near
$v=0$ in the second version of the integral.
This gives $\lambda \to 2\pi/\beta^2$ and $\epsilon \to 2/\beta$.
The same result comes more simply from realizing that only
short distances contribute in the original 2 dimensional
integral. A similar argument can be used for large
negative $\beta$ to give
$\lambda \to 8 e^{\beta/\sqrt{2}}/\beta^2$ and
$\epsilon \to 1/\sqrt{2} - 2/|\beta|$.

\medskip
For the traditional TSP on a unit square domain, the quenched
result is different from the annealed approximation and there
is little hope of a general analytic solution. Particular values,
such as the high temperature $\beta=0$ value may be evaluated (given 
patience with a 4-D integral). In a later section we derive
expansions for large $|\beta|$.

\begin{figure}
\centering\epsfig{file=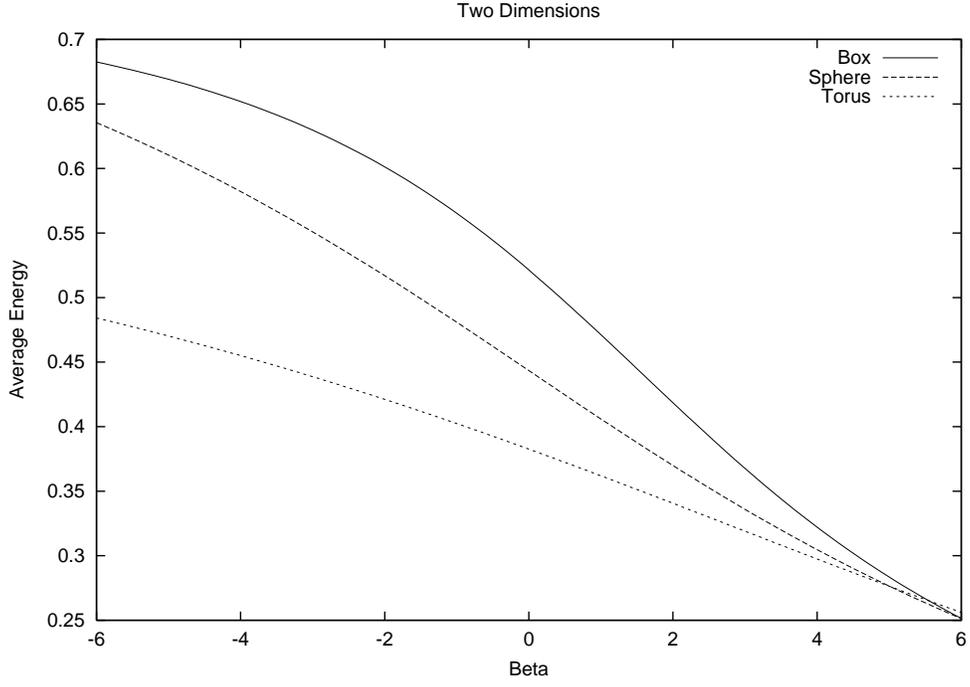,width=0.5\linewidth,angle=270}
\caption{Theoretical prediction for the average energy $\epsilon$ for 
two dimensional TSP on a sphere, box and torus as a function 
of $\beta$.
Negative $\beta$
corresponds to the Maximum TSP problem.}
\label{fig:2DTSP}
\end{figure}

Figure (\ref{fig:2DTSP}) shows the average energy for 
each of the three domains.
For the sphere this is given by (\ref{eqsphere}), for the
torus it is based on numerical integration of (\ref{eqtorus})
and for the box we resort to an iterative solution of the
original quenched equations. The accuracy of this iterative technique is
confirmed by reproducing the results for the other domains.
In all these cases we have also performed Monte Carlo simulations
and obtain excellent agreement with the theory.
At large positive $\beta$ the topology starts to become 
unimportant and each domain has energy $\sim 2/\beta$ as 
expected for a two dimensional TSP.

\subsection{Descent model}
In the case of the descent model, $V(x)$ is clearly not an even function of 
$x$,  and the general approach presented above does not apply, however 
the value of $\lambda_q$ as a functional of $\mu$ can be explicitly computed.
Despite the fact that we know the annealed approximation produces the
correct result we shall pursue our method in this case as it is
rather instructive to do so.

Defining $R(x) = \int_0^x \exp(-\mu(y)/2) f_{R}^{(q)}(y) dy$ in 
 Eq. (\ref{eqfq}) we find
\begin{equation}
\exp(-\mu) {dR\over dx} = \lambda_q^{-1} \left[ (1- \exp(-\beta)) R(x) +
\exp(-\beta) R(1)\right].
\end{equation}
Now we define $y(x) = \int_0^x dy \exp(-\mu(y))$ to obtain
\begin{equation}
{dR\over dy} = \lambda_q^{-1} \left[ (1- \exp(-\beta)) R(y) + \exp(-\beta) R(y(1))
\right].
\end{equation}
This can be solved giving
\begin{equation}
(\exp(\beta)-1){ R(y(x))\over R(y(1))} + 1  = \exp\left(
\lambda_q^{-1} y(x) (1-\exp(-\beta)\right)
\end{equation}
Now setting $x=1$ in the above gives
\begin{equation}
\exp(\beta) = \exp\left(\lambda_q^{-1} y(1) (1-\exp(-\beta))\right),
\end{equation}
which yields
\begin{equation}
\lambda_q = {1-\exp(-\beta)\over \beta } \ \int_0^1 dy 
\exp(-\mu(y)) .
\end{equation}
We thus find that
\begin{equation}
-{\delta \ln(\lambda_q) \over \delta \mu(x)} = {\exp(-\mu(x)) \over 
\int_0^1 dy \exp(-\mu(y)) }.
\end{equation}
A solution (there is again a family related by a constant factor 
giving the same action) to the saddle point equation is
\begin{equation}
\exp(-\mu(x)/2) = \sqrt{p_q(x)}.
\end{equation}
This now yields 
\begin{equation}
{\beta F_N\over N} = -\ln(1-\exp(-\beta)) + \ln(\beta)
+{\rm terms\ independent\ of\ }\beta , 
\end{equation}
This is the same result as the annealed calculation of the 
precedent section as expected.

\subsection{Other one dimensional potentials}
\begin{figure}
\epsfxsize=0.5\hsize \epsfbox{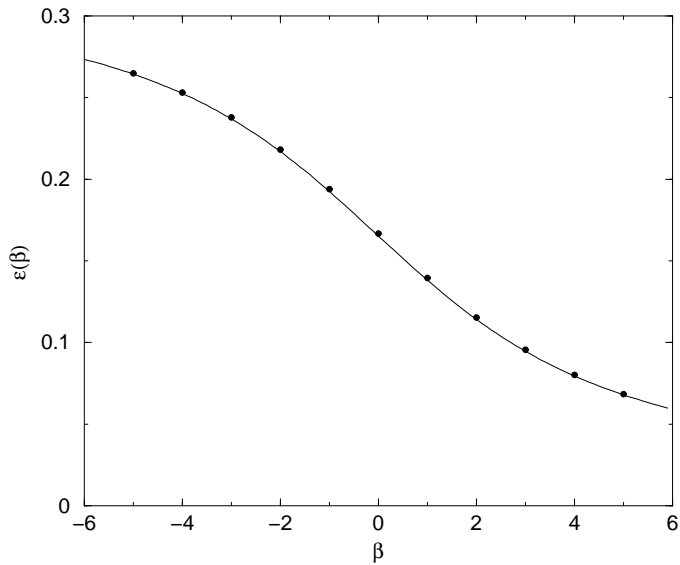}
\caption{Predicted  average energy for potential $V(x) = x^2$ 
as a function  of $\beta$ (solid line) against value measured from 
Monte Carlo simulations
(filled circles)}
\label{fig2}
\end{figure}

\begin{figure}
\epsfxsize=0.5\hsize \epsfbox{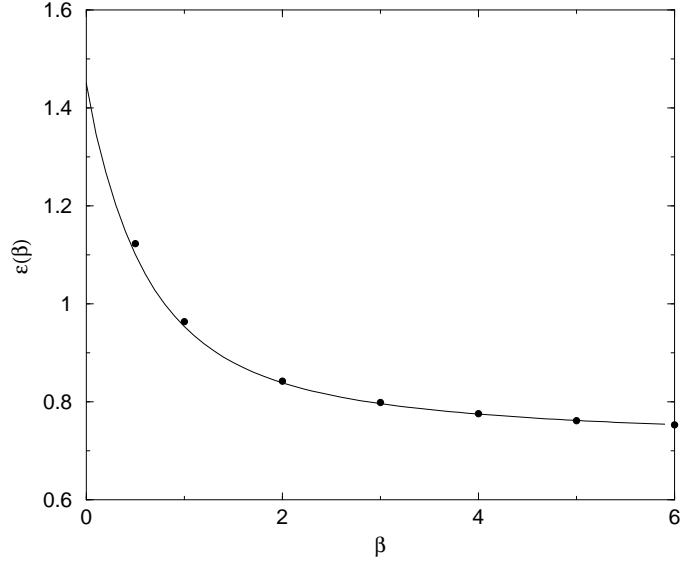}
\caption{Predicted  average energy for potential $V(x) = -\ln(|x|)$ as 
a function  of $\beta$ (solid line) against value measured from 
Monte Carlo simulations
(filled circles)}
\label{fig3}
\end{figure}

As an additional numerical verification of our method we 
have considered the potentials
$V(x) = x^2$ and also $V(x) = -\ln(|x|)$ on the line domain
${\cal D} = [0,1]$.
The latter potential was only considered at 
positive temperature as it is ill defined at negative temperature. The 
comparison of the predictions of our method against results obtained
from Monte Carlo simulations (carried out with the same protocols as for the 
TSP case) are shown in Figs. (\ref{fig2}) and (\ref{fig3}). 
The agreement is again excellent.   

\section{General zero temperature behavior in one dimension}

In some one dimensional cases we may analyze the low temperature 
behavior of Eq. (\ref{eqsff}) and extract the low temperature energy of 
the system analytically. We write $s(x) = \exp(-\beta w(x))t(x)$ and
thus Eq. (\ref{eqsff}) becomes
\begin{equation}
\exp\left(-\beta w(x)\right)t(x) = 
\int dy\ {\exp\left( -\beta V(x-y) +\beta w(y) \right) \over t(y)}.
\label{eqww}
\end{equation}
We now assume that $\ln(t(x))/\beta \to 0$ as $\beta \to \infty$ which
permits us to evaluate the integral on the right hand side of
Eq ({\ref{eqww}) via the saddle point method:
\begin{equation}
\exp(-\beta w(x))t(x) \approx  \exp\left(-\beta r(x)\right)
\int_{-\infty} ^\infty  d\zeta \exp\left[-{\beta\zeta^2\over 2}
\left( V''(x-x^*(x)) - w''(x^*(x))
\right)\right],
\end{equation}
where 
\begin{equation}
 r(x) = V(x-x^*(x)) - w(x^*(x)) = \min_{y\in(0,1)}\left\{ V(x-y) - w(y)
\right\},
\end{equation}
and the point $x^*(x)$ is simply the point about which the 
action in the saddle point is minimal, the fluctuations being integrated 
about this point. It is at this point tempting to suggest a tentative 
physical interpretation of $x^*(x)$ as the optimal point that the 
polymer jumps to if its current position is $x$, that is to say if the monomer
$i$ is at $x$ then the optimal position for monomer $i+1$ is at $x^*(x)$. 
   
The function $w(x)$ is thus determined by $r(x) = w(x)$ {\em i.e}
\begin{equation}
  w(x) = \min_{y\in(0,1)}\left\{ V(x-y) - w(y)\right\}
\label{eqw}
\end{equation}
Intriguing we will see that by effectively guessing some (local) 
heuristics for $x^*(x)$ we will be  able to obtain some solutions to 
Eq. (\ref{eqw}). Putting all this together we obtain the equation for $t$
\begin{equation}
t(x) t(x^*(x)) = \left( {2 \pi \over \beta (V''(x-x^*(x)) - w''(x^*(x))) }
\right)^{1\over 2},
\label{eqt}
\end{equation}
this only being valid if $V''(x-x^*(x)) - w''(x^*(x)) >0$ 
on all but a set of measure zero and when, again on all but a set of
 measure zero,  the minimizing  point occurs within the domain $[0,1]$.
To simplify our analysis we shift the domain $[0,1]$ to the domain $[-{1\over 2}, -{1\over 2}]$, by symmetry we now expect that $w(x) = w(-x)$ and 
$t(x) = t(-x)$. The shifted equation for $w$ is simply
\begin{equation}
  w(x) = \min_{y\in(-{1\over 2},{1\over 2})}\left\{ V(x-y) - w(y)\right\}
\end{equation} 
The energy is now given by
\begin{equation}
\epsilon \approx 2 \int_{-{1\over 2}}^{1\over 2} dx\ w(x)
-2 \ {\partial \over \partial \beta}\int_{-{1\over 2}}^{{1\over 2}} dx \ 
\ln(t(x))
\label{eqelowt}
\end{equation} 
If indeed $x^*(x)$ is the point which is the optimal to jump to from $x$
we expect a one to one correspondence between $x$ and $x^*(x)$ in order 
to generate a uniform annealed distribution. If this is indeed 
the case, then  for any function $F$ on $[0,1]$ we will have
\begin{equation}
\int dx \ F(x) = \int dx \ F(x^*(x))
\label{shift}
\end{equation}
Using Eq. (\ref{shift}) and Eqs. (\ref{eqelowt}) and (\ref{eqt}) we find
\begin{eqnarray}
\epsilon &\approx & 2 \int_{-{1\over 2}}^{1\over 2} dx\ w(x) - \ {\partial \over \partial \beta}\int_{-{1\over 2}}^{1\over 2} dx
\  \ln(t(x)t(x^*(x))) \nonumber \\
&=& 2 \int_{-{1\over 2}}^{1\over 2} dx\ w(x) + {1\over 2\beta}
\label{eqleg}
\end{eqnarray}
Thus, given that the conditions stated above all hold, the correction
to the zero temperature energy at low temperatures takes a remarkably
universal form. 

We first consider the case where $V$ is a purely  attractive potential with 
a minimum at $x=0$. Taking the idea that $x^*(x)$ is the optimal jump
from the point $x$ we expect $x^*(x) = x$. This will imply from
Eq. (\ref{eqw}) that $w(x) = V(0)/2$. This solution can be seen to work when
plugged back into  $Eq. (\ref{eqw})$ when $V'(0) = 0$ and $V''(0) >0$. 
We thus find from Eq. (\ref{eqleg}) that 
\begin{equation}
\epsilon \approx V(0) +{1\over 2\beta}
\end{equation}

We now consider the case where the potential is everywhere repulsive. 
On the interval $[-{1\over 2}, {1\over 2}]$ the greedy algorithm 
described earlier amounts to making the choice $x^*(x) = -x$. This choice
implies that
\begin{equation}
w(x) = V(2x) - w(x)
\end{equation}
and for the solution to be valid we must have that
\begin{equation}
-V'(2x) - w'(-x) = 0.
\end{equation}
If $w(x)=w(-x)$ then we have $w'(x) = -w'(-x)$ and so the above implies that
$w(x) = V(2x)/2$. For this solution to be valid we must have that $V''(2x) <0$
and thus it only holds for concave potentials. 

The low temperature energy is thus given by
\begin{equation}
\epsilon = \int dx\ V(x) + {1\over 2 \beta}.
\end{equation}
The ground state energy is clearly that given by the greedy algorithm
\begin{equation}
\epsilon_{GA} = \int dx\  V(x).
\label{egreedy}
\end{equation} 

Another heuristic for finding the maximal path in the case of repulsive 
potentials is to take a jump size of constant size $\Delta$. When there are no
minima the best value for $\Delta$ is $1/2$, this strategy clearly the 
minimizes the energy at each jump subject to  the constraint that it must 
be possible from  any position $x$. One simply
adds a very small noise $\eta$ to each jump of $1/2$ to generate the required
uniform distribution of monomers on $[0,1]$. We call this the half jump 
algorithm and it is clearly not at all greedy !
We thus take $x^*(x)$ so that $|x -x^*(x)|= 1/2$. When  $x>0$ this
implies $x^*(x) = x- 1/2$ and hence
\begin{equation}
w(x) = V({1\over 2}) - w(x- {1\over 2})
\end{equation}
The explicit solution to this equation is $w(x) = a |x| + b$, where
\begin{equation}
b = V({1\over 2}) -{a\over 2} -b,
\end{equation}
and the condition to 
have a minimum implies that 
\begin{equation}
V'({1\over 2}) -a = 0,
\end{equation}
and $V''(1/2) > 0$. Thus the function $V$ cannot be  concave 
near $x=1/2$ and 
\begin{equation}
b = {1\over 2}\left( V({1\over 2}) -{1\over 2}V'({1\over 2})\right).
\end{equation}
As expected this solution gives
\begin{equation}
\epsilon =  V({1\over 2}) + {1\over 2 \beta}
\end{equation}
which is obviously the ground state energy  given by the half-jump 
algorithm energy.

A potential where the above solution is possible is $V(x)
=-\ln(|x|)$. Numerical solution of Eq. (\ref{eqsff}) at low 
temperatures converges to the solution found above. The 
predicted value of the ground state energy is $\epsilon_{GS} = \ln(2)$, this 
value is compatible with the Monte Carlo simulations for this potential
shown in Fig. (\ref{fig3}). Clearly the greedy algorithm is a bad strategy
for the potential $V(x) = -\ln(|x|)$, this is because at the end it must 
link points very close to each other situated near $x=1/2$, thus 
giving a very large contribution to the energy at the end of the algorithm. 
The greedy algorithm  gives an energy
$\epsilon = -\int dx \ln(x) = 1$, which is indeed higher than the ground 
state we predict analytically and not compatible with our Monte Carlo 
simulations. When $V''(|x|) > 0$
everywhere in $[0,1]$, Jensen's inequality tells us that
$\langle V(X)\rangle \ge V\langle X\rangle$ for $X$ distributed on 
$[0,1]$; when this distribution is uniform this implies that
$\epsilon_{GA} > \epsilon_{HA}$ and hence the  half jump  algorithm is
the most efficient. In the case where the potential
is concave the greedy algorithm is the most efficient. 

We note that the case of the Maximum TSP is an intermediate case where 
$V''(x) = 0$ and in this case $ \epsilon_{GA}= \epsilon_{HA}$ and the 
forms of $u(x)$ in these two cases coincide. As a check of this 
asymptotic analysis we have numerically solved 
Eq. (\ref{eqsff}) for the potential $V(x) = -|x|$ at $\beta = 30$.
Shown in Fig. (\ref{fig4}) is $w^*(x) = -\ln(s(x))/\beta$
where $s(x)$ is the numerical solution of  Eq. (\ref{eqsff})at $\beta = 30$.
Also shown is the predicted zero temperature limit of $w^*$. The agreement
improves on increasing $\beta$ but limitations of numerical accuracy 
are attained if $\beta$ is taken to be too large.

\begin{figure}
\epsfxsize=0.5\hsize \epsfbox{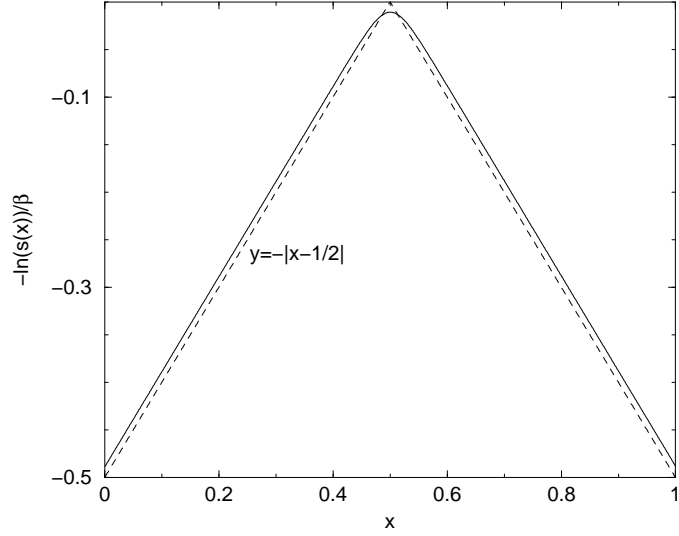}
\caption{Value of $-\ln(s(x))/\beta$ obtained by numerically solving
Eq. (\ref{eqsff}) at $\beta = 30$ (solid line) for the 
potential $V(x) - -|x|$. Also is shown the zero 
temperature prediction for this function (dashed line).
}
\label{fig4}
\end{figure}

The finite temperature
corrections for both the TSP and Maximum TSP are different to those of the
cases considered thus far as $V''(x) = 0$ for $x\neq 0$. At positive 
temperature it is clear that $w(x) = 0$, as the saddle point is at 
$x^*(x) =x$. Here, because $V''(x) =0$, we do not expand the the terms in the 
exponential of the  integral about $y=0$  but we do carry out the 
expansion of the term $1/t(y)$, we thus write
\begin{equation}
t^2(x) \approx \int_{-{1\over 2}}^{1\over 2} dx\ \exp\left(-\beta|x-y|\right)
= {1\over \beta}\left( 2 - \exp(-{\beta\over 2}-\beta x) - 
\exp(-{\beta \over 2} + \beta x)\right)
\end{equation}    
As $\beta \to \infty$ we have
\begin{eqnarray}
\int dx   \ln(s(x)) &\approx& \int_0^{1\over 2} dx \ln\left( 1- {1\over 2} 
\exp(-\beta x)\right) + {1\over 2}\ln\left( {2\over \beta}\right) \nonumber \\
&\approx& {1\over 2\beta}\left[-{\pi^2\over 6} + \ln^2(2)\right] + {1\over 2}\ln\left( {2\over \beta}\right)
\end{eqnarray}
This yields
\begin{equation}
\epsilon \approx {1\over \beta} - 
{1\over \beta^2}\left({\pi^2\over 6} - \ln^2(2)\right).
\end{equation}
for the TSP as $\beta \to \infty$. We have checked that this
result agrees with the asymptotics of our numerical solutions.

The analysis for the Maximum TSP is more involved. Differentiating
Eq. (\ref{eqsf}) twice we see that the function 
$s(x)$ obeys
\begin{equation}
s''(x) = \beta^2 s(x) + {2\beta\over s(x)}. 
\end{equation}
We now make the substitution $s(x) = \exp(\beta|x|)t(x)$ to find
\begin{equation}
t''(x) +2\beta {\rm sgn} (x)t'(x) = -2\beta\delta(x) t(0) + 
{2\beta\exp(-2\beta|x|)\over t(x)}.
\end{equation}
Now in the limit $\beta \to \infty$ we have $2\beta\exp(-2\beta|x|) 
\approx \delta(x)$. For large $\beta$ we thus have
\begin{equation}
t''(x) +2\beta {\rm sgn} (x)t'(x) = \delta(x)\left( -2\beta t(0) + {1\over t(0)}\right).
\end{equation}
The above has solution $t(x) = A + B\exp(-2\beta |x|)$ and the jump
conditions at the origin yield the relation
\begin{equation}
A^2 - B^2 = {1\over 2\beta}.
\end{equation}
An extra relation between $A$ and $B$ is found from examining the 
integral equation for $t$ at $x=0$, which in this limit gives:
\begin{equation}
A+B = {1\over A},
\end{equation}
and in the limit of large $\beta$ we find $A\approx B\approx 1/\sqrt{2}$
and we find that the large $\beta$  behavior of the energy is
\begin{equation}
\epsilon \approx -{1\over 2} +{\pi^2\over 6\beta^2}.
\end{equation}

To end our analysis of the low temperature limit
we will consider the Maximum TSP in higher ($d$) dimensions, specifically on
the hypercube $[0,1]^d$. Consider the generalization of the greedy
heuristic.  Here we start on the outermost layer of points in the
hypercube and we join points on this surface to those that are diametrically
opposed. The procedure is then repeated eating away the hypercube until
we arrive at the center. Shifting the domain to 
$[-{1\over 2},{1\over 2}]^d$ as before,
this entails  matching the point $x$ with $-x$. This generalized heuristic
was shown to give the optimal path length for $d=2$
\cite{or}. The ground state
energy generated by this generalized greedy heuristic is clearly
\begin{eqnarray}
\epsilon_{GA} &=& 2^d \int_{[0,{1\over 2}]^d} dx\  V(2x) \nonumber \\
&=&- 2^{d+1} \int_{[0,{1\over 2}]^d} dx\  |x|.
\end{eqnarray}
This general formula gives $\epsilon_{GA} = -0.5,\ ,-0.765196,\ ,-0.960592$
in one, two and three dimensions respectively. 
The solution $w(x) = -x$ is in fact a solution to Eq. (\ref{eqw})
and thus gives these ground state energies. This can be easily verified
as we note that the function
\begin{equation}
h(x) = |y| - |x-y|
\end{equation}
is bounded as
\begin{equation} 
h(x) \geq - |x|
\end{equation} 
by the triangle inequality. The bound is achieved at $y=-x$, confirming 
that $w(x) = -x$ is indeed a solution. We note that this solution exists
in any domain $\cal D$ (centered at the origin) satisfying the property that  
if $x\in \cal D$ then $-x\in \cal D$.

\section{Conclusions}
We have discussed the statistical mechanics of models whose phase space
is the set of permutations of $N$ objects characterized by quenched positions 
${\bf r}_i$. The Hamiltonians are functions of the 
neighboring elements in the sequence, and thus a given sequence can
be interpreted as the energy of a polymer ring or closed random walk which 
visits all points in the quenched distribution once. 
We analyzed the cases
corresponding to several well studied problems including 
the traveling salesman problem the descent problem
and the quenched Rouse model.
 
The annealed approximation was first considered and 
illustrated for some one-dimensional cases.
For the TSP on a ring and the descent problem this 
annealed approximation gives the correct quenched result.
For the descent model, this is because the effective potential between
neighboring monomers, when the system is viewed as a polymer with
interactions between consecutive monomers, is scale free and independent of
the quenched distribution of the random points. 
For the TSP on a ring, the reason is less clear, but agrees with
expectations from replica studies of the independent link approximation
and continues to hold for symmetric closed domains in higher dimensions.
However in general 
we expect the  annealed approximation to fail at all but infinite 
temperature. This is because the points ${\bf r}_i$ are allowed to evolve 
dynamically to lower the free energy of the system and the
resulting thermodynamic distribution will not be the same as their original
quenched distribution. 

We then showed how the quenched calculation 
could be carried out and confirmed its predictions for both
one and two dimensional TSP examples with Monte Carlo 
simulations. Physically the method we introduced corresponds to imposing a 
fictitious site dependent chemical potential on the distribution of a set 
of dynamical variables ${\bf r}_i$ in the presence of the original interaction 
Hamiltonian. This chemical potential is then chosen to ensure that the 
annealed distribution of the positions of these dynamical ${\bf r}_i$, 
denoted in 
this paper by $p_a({\bf r})$, is the same as the quenched distribution of the 
quenched random variables ${\bf r}_i^{(q)}$ denoted by $p_q({\bf r})$. 
The method is exact
in the thermodynamic limit (corresponding
to high density where the length of the interval is held constant)
for any quenched distribution $p_q({\bf r})$ and interaction potential
$V({\bf r})$. 
On of the most intriguing observations made here is that in the 
annealed approximation we are lead to consider a  linear eigenvalue 
problem to solve the thermodynamics as we use a 
transfer matrix approach, however in the quenched calculation we
are lead to consider a non-linear integral equation. Although we 
managed to avoid the sometimes rather opaque replica method
in our treatment, it would be interesting to see if the appearance
of the non-linear eigenvalue equation could be interpreted or re-derived
within the replica formalism. The results of our calculations were then
confirmed by comparing them with Monte Carlo simulations in a variety of 
models. In the cases we have considered so far we have seen no evidence
for any phase transition on lowering the system's temperature. It is
possible however that in higher dimensions and with certain interaction
potentials $V$ that a phase transition does occur.  
We recall that the 
a directed polymer in dimensions greater than two exhibits a finite 
temperature phase transition \cite{dego}.

Particular attention was paid to the zero temperature limit where 
Eq. (\ref{eqw}) needs to be solved. A number of solutions were found 
which, although we did not prove uniqueness, are compatible with our
numerical simulations and also a rigorous result for the Maximum TSP
in two dimensions. The solution of Eq. (\ref{eqw}) was carried out by 
trying out different heuristics to construct the optimal path, it is
possible that in more complex situations the method used here could
serve as a useful indicator for such constructions.

Finally the idea of treating quenched variables as
effectively annealed variables and then adjusting their Boltzmann  
weight in order to recover self consistently the original quenched 
distribution may prove useful, either as an exact or approximate method 
in other problems involving quenched disorder. Indeed Morita's 
pioneering work used this idea in an approximate context, here we
have shown that the procedure can be carried out exactly
for this type of permutation based combinatorial optimization problem, 
as is also the case for some one dimensional spin  models \cite{kuhn}.

\vskip 0,5 truecm

\noindent{\bf Acknowledgement:} We would like to thank N. Read for drawing our 
attention to reference \cite{or}. D.S.D. would like to acknowledge that part
of this work was carried out at the KITP, UCSB and was supported in part by the
National Science Foundation under Grant No. PHY99-0794.
 
\pagestyle{plain}

\end{document}